\documentclass[english,aps,twocolumn,pra]{revtex4}
\usepackage[latin9]{inputenc}
\usepackage{amsmath}
\usepackage{graphicx}
\usepackage{esint}

\makeatletter
\@ifundefined{textcolor}{}
{%
 \definecolor{BLACK}{gray}{0}
 \definecolor{WHITE}{gray}{1}
 \definecolor{RED}{rgb}{1,0,0}
 \definecolor{GREEN}{rgb}{0,1,0}
 \definecolor{BLUE}{rgb}{0,0,1}
 \definecolor{CYAN}{cmyk}{1,0,0,0}
 \definecolor{MAGENTA}{cmyk}{0,1,0,0}
 \definecolor{YELLOW}{cmyk}{0,0,1,0}
}

\makeatother

\usepackage{babel}
\begin{document}

\global\long\def\r{\left(\mathbf{r}\right)}

\global\long\def\rp{\left(\mathbf{r}'\right)}

\global\long\def\rt{\left(\mathbf{r}\tau\right)}

\global\long\def\rtp{\left(\mathbf{r}\tau'\right)}

\global\long\def\rpt{\left(\mathbf{r}'\tau\right)}

\global\long\def\rptp{\left(\mathbf{r}'\tau'\right)}

\global\long\def\Sh{\hat{\mathbf{S}}}

\global\long\def\rtrptp{\left(\mathbf{r}\tau;\mathbf{r}'\tau'\right)}

\global\long\def\dSh{\cdot\hat{\mathbf{S}}}

\global\long\def\n{\mathbf{n}}

\global\long\def\np{\dot{\mathbf{n}}}

\global\long\def\nnp{\mathbf{n}\cdot\dot{\mathbf{n}}}

\global\long\def\nknp{\mathbf{n}\times\dot{\mathbf{n}}}

\global\long\def\nh{\hat{n}}

\global\long\def\a{\alpha}

\global\long\def\b{\beta}

\global\long\def\s{\sigma}

\global\long\def\D{\Delta}

\global\long\def\sp{\sigma'}

\global\long\def\sb{\sigma"}

\global\long\def\Lb{{\bf \Lambda}}

\global\long\def\wm{\omega_{m}}

\global\long\def\wn{\omega_{n}}

\global\long\def\wl{\omega_{\ell}}

\global\long\def\wmp{\omega_{m'}}

\global\long\def\wnp{\omega_{n'}}

\global\long\def\kwm{\left(\mathbf{k}\omega_{m}\right)}

\global\long\def\uz{U_{0}}

\global\long\def\ud{U_{2}}

\global\long\def\d{\partial}

\global\long\def\e{\epsilon}

\global\long\def\mcS{\mathcal{S}}

\global\long\def\mcD{\mathcal{D}}

\global\long\def\mcH{\mathcal{H}}

\global\long\def\Ek{E_{\mathbf{k}}}

\global\long\def\zz{\left(\mathbf{0}0\right)}


\title{Temperature-dependent excitation spectra of ultra-cold bosons in
optical lattices}

\author{T. A. Zaleski\footnote{Corresponding author. Tel.: +48 713435021; fax: +48 713441029.
\textit{E-mail address}: t.zaleski@int.pan.wroc.pl (T.Zaleski).}, T. K. Kope\'{c}}

\affiliation{Institute of Low Temperature and Structure Research, Polish Academy
of Sciences, POB 1410, 50-950 Wroc\l{}aw 2, Poland}
\begin{abstract}
Trapping ultra-cold atoms in optical lattices provides a unique environment
for investigating quantum phase transitions between strongly correlated
superfluid and Mott insulator phases. However, one of the major complications
in the analysis of experiments are criteria for identifying the superfluid
phase. Sharp features occurring while entering ordered state have
been recognized as a signature of superfluidity. Here, we show that
sharp peaks are not necessarily a reliable diagnostic of phase coherence
in these systems. Using the combined Bogoliubov method and the quantum
rotor approach for phase variables, we calculate the momentum and
energy-resolved single-particle spectral function at arbitrary temperature
and its shape in the presence of the superfluid phase. We find that
in the two-dimensional system even even at $T>0$, where condensate
fraction vanishes, the remnants of the sharp coherence peak are present.
In contrast, such a feature is not observed for the bosons loaded
in the three-dimensional lattice.
\end{abstract}



\maketitle

\section{Introduction}

Loading of ultra-cold atoms into an optical lattice allows to study
an ``artificial solid'' that is undisturbed by any imperfections
present in real materials \cite{optical_lattices}. It also allows
to investigate a regime of strong correlations, when the energy of
interactions between particles overcomes their kinetic energy \cite{quantum_many_body,strong_correlations}.
For example, in the realm of condensed matter physics, they manifest,
e.g., in optical excitation spectra, which shows spectral weight transfers
connecting different energy scales \cite{kopec_1}. In the discussion
of photoemission on solids, and in particular on the correlated electron
systems, the most powerful and commonly used approach is based on
the Green's--function formalism. In this context, the propagation
of a single electron in a many-body system is described by the time-ordered
one-electron Green's propagator. A common approach in describing strong
electron correlations is based on consideration of the Hubbard model
\cite{hubbard_1,hubbard_2}. In the context of bosonic systems, these
properties can be also studied in systems of ultra-cold gases in optical
lattices via methods based on a response to scattering of photons:
radio-frequency spectroscopy \cite{exp2}, Raman spectroscopy \cite{exp3}
and Bragg spectroscopy \cite{exp4,exp7,exp7b}. These experiments
reveal the band structure of these systems, which can be used to compare
the quantum gas phases with their condensed-matter counterparts. One
key piece of evidence for the Mott insulator phase transition is the
loss of global phase coherence of the matter wave function. However,
there are many possible sources of phase decoherence in these systems.
Substantial decoherence can be induced by quantum or thermal depletion
of the condensate, so loss of coherence is not a proof that the system
resides in the Mott insulator ground state. Although the initial system
can be prepared at a relatively low temperature, the ensuing system
after ramp-up of the lattice has a temperature which is usually higher
due to adiabatic and other heating mechanisms. Recent experiments
have reported temperatures on the order of $k_{B}T\sim0.9t$ where
$t$, the hopping parameter, measures the kinetic energy of bosons
\cite{temperature}. At such temperatures, the effects of excited
states become important, motivating investigations of the the \textit{finite
}temperature phase diagrams, showing the interplay between quantum
and thermal fluctuations. 

In the present paper we study combined effects of a confining lattice
potential and finite temperature on the excitation spectra of the
Bose-Hubbard model in two and three dimensions. The relevant part
of the low-energy excited states involve one or two particles of the
many-body ground-state and they are respectively connected to the
momentum $\mathbf{k}$ and energy $\omega$-dependent one-particle
spectral function $A\left(\mathbf{k}\omega\right)$, which describes
the probability of removing a particle from the system and creating
an excitation. This is precisely the quantity, which we want to study
in the present paper. To this end, w perform calculations of the one-particle
spectral function for the Bose-Hubbard model on the square and cubic
lattice using a recently developed the quantum U(1) rotor approach
\cite{zaleski_spectral}. The collective variables for phase are isolated
in the form of the space--time fluctuating phase field governed by
the U(1) symmetry. As a result interacting particles appear as a composite
objects consisting of bare bosons with attached U(1) gauge fields.
This allows us to write the boson Green's function in the space-time
domain as the product of the U(1) phase propagator and the bare boson
correlation function. The problem of calculating the spectral line
shapes now becomes one of determining the convolution of phase and
bare boson Green's functions. 

The plan of the paper is as follows: first, in Section II we introduce
the microscopic Bose-Hubbard model that is an accepted description
of strongly interacting bosons in an optical lattice. Furthermore,
in Sections III and IV we briefly present the used method and arrive
at the expressions for the one-particle spectral function. This allows
us to present in Section V the resulting excitation spectra of bosons
in two- and three-dimensional optical lattice and discuss the influence
of the temperature on the shape of spectral lines and the presence
of superfluid phase. Finally, we conclude and summarize our results
in the final Section VI.

\section{Model and Way of Treatment}

\subsection{Hamiltonian}

Ultra-cold bosonic atoms in move an in optical lattice within a tight-binding
scheme. They are well described by the microscopic Bose-Hubbard Hamiltonian
\cite{jaksch}. The atoms are neutral, so their interactions are realized
via collision. This results in an on-site repulsion, which is responsible
for inter-bosonic correlations. As a result, we consider a second
quantized, bosonic Hubbard Hamiltonian in the form \cite{bemodel1,bemodel2}:

\begin{eqnarray}
\mathcal{H} & = & -t\sum_{\left\langle \mathbf{r},\mathbf{r}'\right\rangle }\left[a^{\dagger}\left(\mathbf{r}\right)a\left(\mathbf{r}'\right)+a^{\dagger}\left(\mathbf{r}'\right)a\left(\mathbf{r}\right)\right]\nonumber \\
 &  & +\frac{U}{2}\sum_{\mathbf{r}}n^{2}\r-\overline{\mu}\sum_{\mathbf{r}}n\left(\mathbf{r}\right).\label{eq:mainham}
\end{eqnarray}
The hopping of bosons between neighboring lattice sites $\mathbf{r}$
and $\mathbf{r}'$ is described by the first term with tunneling element
$t$ and operators $a^{\dagger}\left(\mathbf{r}\right)$ and $a\left(\mathbf{r}'\right)$,
which create and annihilate bosons (thus, $n\r=a^{\dagger}\left(\mathbf{r}\right)a\r$
is a boson number operator). Summation $\left\langle \mathbf{r},\mathbf{r}'\right\rangle $
runs over nearest neighbors of a regular two- (2D) or three-dimensional
(3D) lattice. The on-site repulsion depending on the number of atoms
occupying every lattice site is given by $U$ term. Finally, the chemical
potential $\mu$ (with $\overline{\mu}=\mu+U/2$) controls the total
number of bosons introduced to the lattice (with $N$ being the total
number of lattice sites). The Hamiltonian and its modifications have
been widely studied within the recent years using both analytical
\cite{key-1,key-4,key-5} and numerical \cite{key-9,key-13} methods.

\subsection{Path integral formulation}

The statistical sum of the system can be expressed in a path integral
form with use of complex fields, $a\left(\mathbf{r}\tau\right)$ depending
on the imaginary time $0\le\tau\le\beta\equiv1/k_{B}T$ (with $T$
being the temperature) 
\begin{equation}
Z=\int\left[\mathcal{D}\overline{a}\mathcal{D}a\right]e^{-\mathcal{S}\left[\overline{a},a\right]},
\end{equation}
where the action $\mathcal{S}$ is given by
\begin{equation}
\mathcal{S}\left[\overline{a},a\right]=\int_{0}^{\beta}d\tau\left[\mathcal{H}\left(\tau\right)+\sum_{\mathbf{r}}\overline{a}\left(\mathbf{r}\tau\right)\frac{\partial}{\partial\tau}a\left(\mathbf{r}\tau\right)\right].
\end{equation}
The complex fields $a\rt$ satisfy the periodic condition $a\left(\mathbf{r}\tau\right)=a\left(\mathbf{r}\tau+\beta\right)$.
We use a recently proposed method based on a combination of Bogoliubov
and quantum rotor approaches. Our way of treatment of the model has
been presented in details in Ref. \cite{zaleski_tof}, so here we
restrict ourselves to listing the main steps only.

\section{Single-particle correlation function}

In order to determine spectral properties of the model, it is necessary
to determine the one-particle Green's function of the system defined
as: 
\begin{equation}
G\left(\mathbf{r}-\mathbf{r}';\tau-\tau'\right)=\left\langle a\rt\overline{a}\rptp\right\rangle _{a},\label{eq:GreenDef}
\end{equation}
where statistical averaging
\begin{equation}
\left\langle \dots\right\rangle _{a}=\frac{1}{Z}\int\left[\mathcal{D}\overline{a}\mathcal{D}a\right]\dots e^{-\mathcal{S}\left[\overline{a},a\right]}.
\end{equation}
The Green function can be expressed using Fourier transform: 
\begin{equation}
G\left(\mathbf{r};\tau\right)=\frac{1}{\beta N}\sum_{\mathbf{k},m}G\kwm e^{i\mathbf{k}\mathbf{r}+i\omega_{m}\tau},\label{eq:GreenFourier}
\end{equation}
where $\wm=2\pi m/\beta$ is Matsubara frequency with $m$ being an
integer value. The spectral function is defined as the imaginary part
of the Green's function:
\begin{equation}
A\left(\mathbf{k}\omega\right)=2\mathrm{Im}G\left(\mathbf{k};i\omega_{m}\rightarrow\omega+i0^{+}\right)\label{eq:SpectralDef}
\end{equation}
and relation of $A\left(\mathbf{k}\omega\right)$ to the function
given in Eq. \eqref{eq:GreenFourier} is established through the Hilbert
transform:
\begin{equation}
G\left(\mathbf{k}\omega_{m}\right)=-\int_{-\infty}^{+\infty}\frac{d\omega}{2\pi}\frac{A\left(\mathbf{k}\omega\right)}{i\omega_{m}-\omega}.\label{eq:spectral_definition}
\end{equation}
Furthermore, from the Eqs. \eqref{eq:SpectralDef} and \eqref{eq:spectral_definition}
a sum rule appears:
\begin{equation}
\frac{1}{N}\sum_{\mathbf{k}}\int_{-\infty}^{+\infty}d\omega A\left(\mathbf{k}\omega\right)=1,
\end{equation}
which means that the spectral function is normalized. From the spectral
function, one can extract the excitation spectrum and the strength
of the excitation modes.

\section{Introduction of phase variables}

In order to advance calculations of the Green function in Eq. \eqref{eq:GreenDef},
we resort on the quantum rotor approach (see, Ref. \cite{polak})
combined with the Bogoliubov method \cite{bogoliubov} that has been
recently proposed and successfully applied to systems of bosons in
optical lattices. This scenario provides a picture of quasiparticles
and energy excitations in the strong-interaction limit, where the
transition between the superfluid and the Mott state is driven by
phase fluctuations. The essence of the approach is the separation
of the original Bose field into its amplitude and fluctuating phase
that was absent in the original Bogoliubov treatment. As a result,
one arrives at a formalism where the one-particle correlation functions
are treated self-consistently and permit us to investigate a whole
range of phenomena described by the Bose-Hubbard Hamiltonian. Furthermore,
the phase fluctuations are described within the quantum spherical
model \cite{vojta}, which goes beyond the mean-field approximation,
including both quantum and spatial correlations. The method is based
on a local gauge transformation to the new bosonic variables: 
\begin{equation}
a\rt=b\rt e^{i\phi\rt},\label{eq:gauge_transformation}
\end{equation}
which allows extraction of the phase variable $\phi\rt$, whose ordering
naturally describes the superfluid\textendash{}Mott-insulator transition,
and the amplitude $b\rt$, which is related to the superfluid density.
As a result, the original Green's function defined in Eq. \eqref{eq:GreenDef}
takes a form of a product of the phase and bosonic correlators:
\begin{equation}
G\left(\mathbf{r}\tau;\mathbf{r}'\tau\right)=G_{\phi}\left(\mathbf{r}\tau;\mathbf{r}'\tau\right)G_{b}\left(\mathbf{r}\tau;\mathbf{r}'\tau\right),\label{eq:GreenProduct}
\end{equation}
which are defined as:
\begin{eqnarray}
G_{b}\left(\mathbf{r}\tau;\mathbf{r}'\tau\right) & = & \left\langle b\rt\overline{b}\rpt\right\rangle _{b},\nonumber \\
G_{\phi}\left(\mathbf{r}\tau;\mathbf{r}'\tau\right) & = & \left\langle e^{i\phi\rt-i\phi\rpt}\right\rangle _{\phi}.
\end{eqnarray}
In the Fourier space the relation in Eq. \eqref{eq:GreenProduct}
becomes: 

\begin{equation}
G\left(\mathbf{k}\omega_{m}\right)=\frac{1}{\beta N}\sum_{\mathbf{k}'m'}G_{b}\left(\mathbf{k}'\omega_{m'}\right)G_{\phi}\left(\mathbf{k}-\mathbf{k}';\omega_{m}-\omega_{m'}\right),\label{eq:greens_function_ft}
\end{equation}
where the spectral decomposition of each constituent in Eq. \eqref{eq:greens_function_ft}
is given by:
\begin{equation}
G_{x}\left(\mathbf{k}\omega_{m}\right)=-\int_{-\infty}^{+\infty}\frac{d\omega}{2\pi}\frac{A_{x}\left(\mathbf{k}\omega\right)}{i\omega_{m}-\omega},
\end{equation}
with $x=b,\phi$. Furthermore, we separate the bosonic amplitude $b\rt$
into the condensed $b_{0}$ and non-condensed $b_{d}\rt$ part, following
the Bogoliubov approach \cite{bogoliubov}

\begin{equation}
b\rt=b_{0}+b_{d}\rt.
\end{equation}
The superfluid order parameter becomes:
\begin{equation}
\Psi_{B}\equiv\left\langle a\rt\right\rangle =b_{0}\psi_{B},
\end{equation}
where $\psi_{B}$ is the phase order parameter defined as:
\begin{equation}
\psi_{B}=e^{i\phi\rt}.
\end{equation}
As a result, the spectral function corresponding to the original Green's
function in Eq. \eqref{eq:SpectralDef} can be expressed as follows:

\begin{eqnarray}
 &  & A\left(\mathbf{k}\omega\right)=b_{0}^{2}A_{\phi}\left(\mathbf{k}\omega\right)+\psi_{B}^{2}A_{b}\left(\mathbf{k}\omega\right)\nonumber \\
 &  & \,\,-\frac{1}{N}\sum_{\mathbf{k}'}\int_{-\infty}^{+\infty}\frac{d\omega'}{2\pi}A_{b}\left(\mathbf{k}'\omega'\right)A_{\phi}\left(\mathbf{k}-\mathbf{k}';\omega-\omega'\right)\nonumber \\
 &  & \,\,\times\left[\frac{1}{e^{-\beta\omega'}-1}-\frac{1}{e^{\beta\left(\omega-\omega'\right)}-1}\right],\label{eq:spectral_function_ft}
\end{eqnarray}
where the first and the second terms represent macroscopic ordered
states resulting from emergence of bosonic Bogoliubov amplitude $b_{0}$
and phase order parameter $\psi_{B}$, respectively. They will lead
to appearance of sharp coherence peaks in the excitation spectrum
for non-zero values of $b_{0}$ and $\psi_{B}$. The last is the remaining
disordered part of the spectrum. The spectral functions $A_{b}\left(\mathbf{k}\omega\right)$
and $A_{\zeta}\left(\mathbf{k}\omega\right)$ can be written explicitly
following our earlier work (see, Ref. \cite{zaleski_spectral} for
details):
\begin{eqnarray}
A_{b}\left(\mathbf{k}\omega\right) & = & E_{+}\left(\mathbf{k}\right)\delta\left(\omega-E_{\mathbf{k}}\right)-E_{-}\left(\mathbf{k}\right)\delta\left(\omega+E_{\mathbf{k}}\right)\nonumber \\
A_{\phi}\left(\mathbf{k}\omega\right) & = & \frac{\pi}{\Xi\left(\mathbf{k}\right)}\left\{ \delta\left[\frac{\omega}{U}+v\left(\frac{\overline{\mu}}{U}\right)-\Xi\left(\mathbf{k}\right)\right]\right.\nonumber \\
 &  & \left.-\delta\left[\frac{\omega}{U}+v\left(\frac{\overline{\mu}}{U}\right)+\Xi\left(\mathbf{k}\right)\right]\right\} ,\label{eq:spectral_functions_sectors}
\end{eqnarray}
where:
\begin{eqnarray}
 &  & E_{\mathbf{k}}=2\sqrt{t\left(\varepsilon_{0}-\varepsilon_{\mathbf{k}}\right)\left[t\left(\varepsilon_{0}-\varepsilon_{\mathbf{k}}\right)+Ub_{0}^{2}\right]}\nonumber \\
 &  & E_{\pm}\left(\mathbf{k}\right)=2\pi\left[\frac{2t\left(\varepsilon_{0}-\varepsilon_{\mathbf{k}}\right)+b_{0}^{2}U}{2E_{\mathbf{k}}}\pm\frac{1}{2}\right]\nonumber \\
 &  & \Xi\left(\mathbf{k}\right)=\sqrt{\frac{\delta\lambda}{U}+\frac{2t}{U}b_{0}^{2}\left(\varepsilon_{\mathbf{0}}-\varepsilon_{\mathbf{k}}\right)+v^{2}\left(\frac{\overline{\mu}}{U}\right)}.\label{eq:dodatkowe}
\end{eqnarray}
Furthermore, $v\left(x\right)=x-\left[x\right]-\frac{1}{2}$, with
$\left[x\right]$ being the floor function (giving the greatest integer
bigger or equal to $x$) and the Bogoliubov amplitude $b_{0}$ can
be determined \cite{zaleski_tof}: 
\begin{equation}
b_{0}^{2}=\frac{4t}{U}+\frac{\overline{\mu}}{U}.
\end{equation}
Next, $\delta\lambda$ is the parameter, which arrises in the quantum
rotor treatment \cite{polak,zaleski_spectral} that measures ``the
distance from criticality'' in the high-temperature phase ($\delta\lambda=0$
at the critical point and in the ordered, low-temperature phase).
Values of $\delta\lambda$ (in the disordered phase) and the phase
order parameter $\psi_{B}$ (in the ordered phase) and are given by
the equation (see, Ref. \cite{zaleski_spectral}): 
\begin{eqnarray}
1-\psi_{B}^{2} & = & \frac{1}{N}\sum_{\mathbf{k}}\frac{\coth\left\{ \frac{\beta U}{2}\left[\Xi\left(\mathbf{k}\right)-v\left(\frac{\overline{\mu}}{U}\right)\right]\right\} }{4\Xi\left(\mathbf{k}\right)}\nonumber \\
 & + & \frac{1}{N}\sum_{\mathbf{k}}\frac{\coth\left\{ \frac{\beta U}{2}\left[\Xi\left(\mathbf{k}\right)+v\left(\frac{\overline{\mu}}{U}\right)\right]\right\} }{4\Xi\left(\mathbf{k}\right)}.\label{eq:criticalline}
\end{eqnarray}
Finally, the lattice structure factor in Eq. \eqref{eq:dodatkowe}
is defined as
\begin{equation}
\varepsilon_{\mathbf{k}}=\frac{1}{2}\sum_{\mathbf{d}}e^{i\mathbf{k}\mathbf{d}}\label{eq:dispersiondef}
\end{equation}
with the summation running over the lattice vector $\mathbf{d}$,
which specifies range of bosonic hopping. In order to evaluate expressions
in Eqs. \eqref{eq:spectral_function_ft}-\eqref{eq:criticalline}
explicitly we need to specify a lattice using its dispersion relation
$\varepsilon_{\mathbf{k}}$ in Eq. \eqref{eq:dispersiondef}.

\section{Excitation Spectra}

In the present paper we consider two relevant examples of experimentally
realizable optical lattice, namely two-dimensional regular square
lattice with dispersion $\varepsilon_{\mathbf{k}}$ is given by 
\begin{equation}
\varepsilon_{\mathbf{k}}=\cos\left(ak_{x}\right)+\cos\left(ak_{y}\right),
\end{equation}
($a$ denotes the lattice constant) and three-dimensional simple cubic
lattice:
\begin{equation}
\varepsilon_{\mathbf{k}}=\cos\left(ak_{x}\right)+\cos\left(ak_{y}\right)+\cos\left(ak_{z}\right),
\end{equation}
which follows from Eq. \eqref{eq:dispersiondef} with the assumption
that the summation runs over the nearest neighboring lattice sites.
\begin{figure}
\includegraphics[scale=0.3]{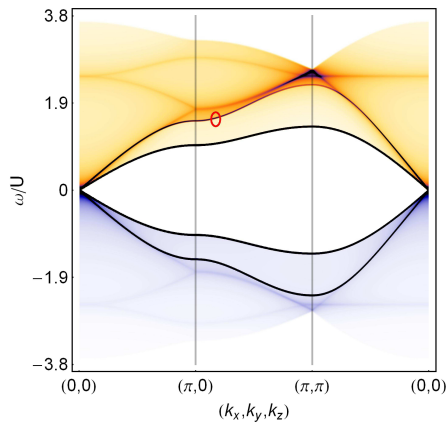}

\caption{(Color online) Density plot of the spectral function for two-dimensional
system at $k_{B}T/U=0$ and $t/U=0.15$, $\mu/U=0.5$ (darker areas
depicts higher value) along with cuts of absolute value of $A\left(\mathbf{k}\omega\right)$
for selected values of momentum. For $\omega/U>0$, $A\left(\mathbf{k}\omega\right)>0$
(brown), while for $\omega/U<0$,\textbf{ $A\left(\mathbf{k}\omega\right)<0$}
(blue). The red circle depicts position in the momentum space, for
which Fig. \ref{fig:peakmagnification} is plotted.\label{fig:density2d}}
\end{figure}
In order to investigate the excitation spectrum as a function on momentum
and energy, we need to calculate the values of $A\left(\mathbf{k}\omega\right)$
from Eq. \eqref{eq:spectral_function_ft}. The resulting spectra are
presented as density plots, in which darker areas depict larger absolute
values. Since, for bosons $A\left(\mathbf{k}\omega\right)$ takes
both positive and negative values, they are denoted by different colors
on the plots: blue and yellow for $A\left(\mathbf{k}\omega\right)<0$
and $A\left(\mathbf{k}\omega\right)>0,$ respectively. The spectra
are presented along selected cuts of the $\mathbf{k}$-space: $\left(0,0\right)\rightarrow\left(\pi,0\right)\rightarrow\left(\pi,\pi\right)\rightarrow\left(0,0\right)$
for two-dimensional system and $\left(0,0,0\right)\rightarrow\left(\pi,0,0\right)\rightarrow\left(\pi,\pi,0\right)\rightarrow\left(\pi,\pi,\pi\right)\rightarrow\left(0,0,0\right)$
-- for three-dimensional one. Plot in Fig. \ref{fig:density2d} presents
spectra of the system in the superfluid phase (for the phase diagram
of Bose-Hubbard system, see Ref. \cite{zaleski_diagram}). The spectrum
is gapless, with two narrow coherence lines: the inner one results
from the presence of non-zero Bogoliubov amplitude, while the outer
reflects existence of the phase-ordered phase. The remainder of the
spectrum is smeared and comes from incoherent (although strongly interacting)
atoms. The increase of interactions between atoms $t/U$ leads to
grow of phase fluctuations, which ultimately destroy the superfluid
phase and drive the system into the Mott insulator. However, as temperature
increases, thermal fluctuations melt away both the superfluid and
Mott-insulating phases, introducing the normal phase \cite{trivedinonzerot}.

\subsection{Two-dimensional system}

\begin{figure}
\includegraphics[scale=0.4]{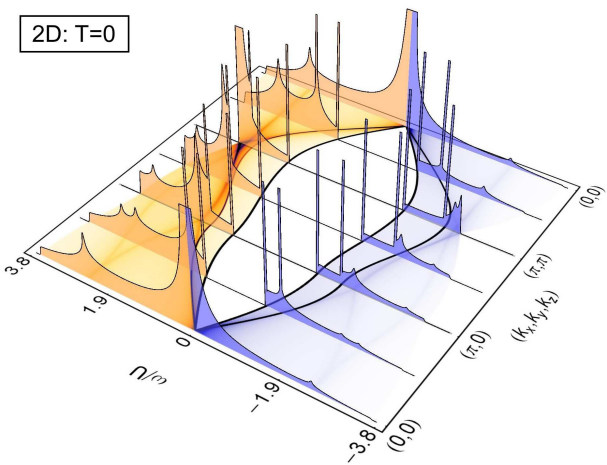}

\caption{(Color online) Density plot of the spectral function for two-dimensional
system at $k_{B}T=0$ and $t/U=0.15$, $\mu/U=0.5$ (darker areas
depicts higher value) along with cuts of absolute value of $A\left(\mathbf{k}\omega\right)$
for selected values of momentum. For $\omega/U>0$, $A\left(\mathbf{k}\omega\right)>0$
(brown), while for $\omega/U<0$,\textbf{ $A\left(\mathbf{k}\omega\right)<0$}
(blue).\label{fig:2dT0}}
\end{figure}

\begin{figure}
\includegraphics[bb=20bp 20bp 592bp 480bp,scale=0.4]{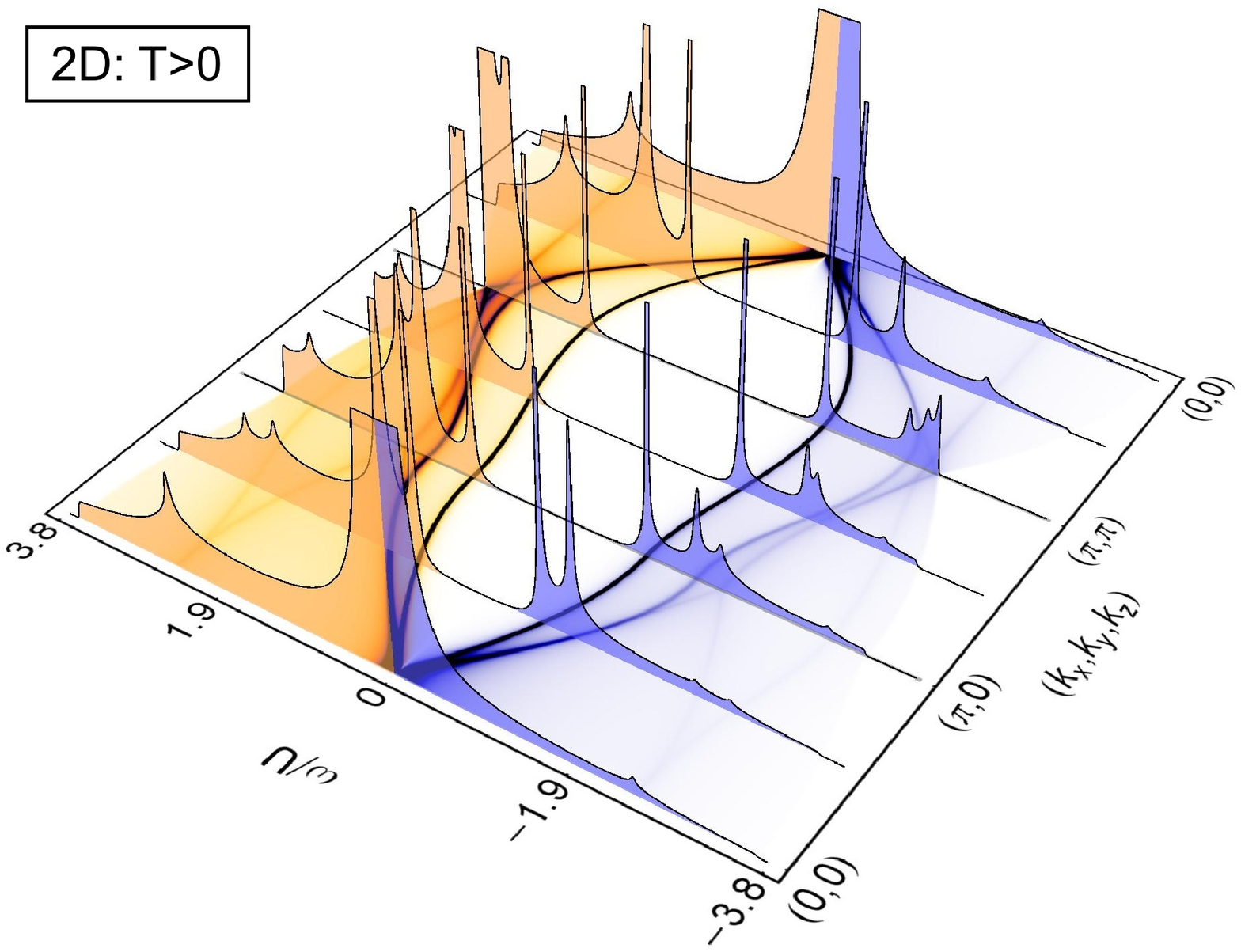}

\caption{(Color online) Density plot of the spectral function for two-dimensional
system at $k_{B}T/U=0.1$ and $t/U=0.15$, $\mu/U=0.5$ (darker areas
depicts higher value) along with cuts of absolute value of $A\left(\mathbf{k}\omega\right)$
for selected values of momentum. For $\omega/U>0$, $A\left(\mathbf{k}\omega\right)>0$
(brown), while for $\omega/U<0$,\textbf{ $A\left(\mathbf{k}\omega\right)<0$}
(blue).\label{fig:2dTh}}
\end{figure}

\begin{figure}
\includegraphics[scale=0.4]{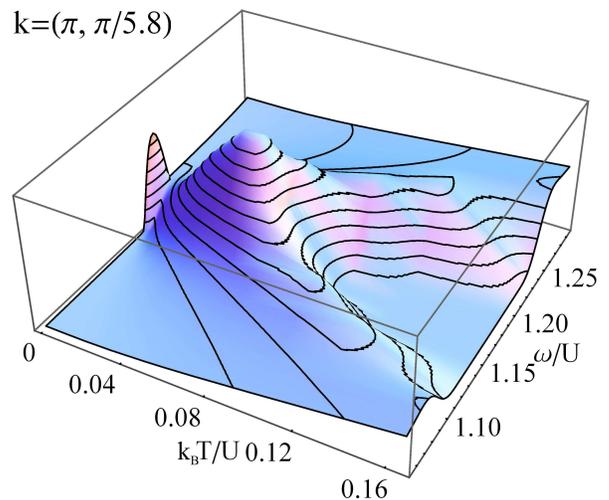}

\caption{(Color online) Behavior of the superfluid coherence peak within spectral
function $A\left(\mathbf{k}\omega\right)$ with increasing temperature
for two-dimensional system at selected momentum $k=\left(\pi,\pi/5.8\right)$,
for $t/U=$ and $\mu/U=0.5$.\label{fig:peakmagnification}}
\end{figure}

In the two-dimensional system the long-range phase order occurs only
at zero temperature (see, Fig. \ref{fig:2dT0}). Any increase of the
temperature destroys the superfluid phase ($\psi_{B}=0$). This results
in immediate disappearance of sharp coherence peak associated with
the phase order parameter, however value of $\delta\lambda$ is small
so the gap at $\mathbf{k}=0$ is not noticeable. Surprisingly, in
higher temperatures the peak seems to be restored, although it is
now a part of the incoherent particles spectrum resulting from convolution
of phase and bosonic spectral functions (see, Fig. \ref{fig:2dTh}).
Finally, further increase of the temperature leads to the peak being
replaced by a dip. The temperature evolution of the peak is presented
in Fig. \ref{fig:peakmagnification} (its coordinates in the momentum
space have been denoted by a red circle in Fig. \ref{fig:density2d}).
As a result, in two-dimensional system there is an evidence of the
quasi long-range phase order in finite temperatures. Another influence
of thermal fluctuations on excitation spectra visible in Fig. \ref{fig:2dTh}
is smearing around the inner edges of bands (for low values of $\left|\omega/U\right|$).
Because the effect is subtle, additional energy cuts have been presented
in Figs. \ref{fig:2dT0}-\ref{fig:2dTh} (and similar figures for
the three-dimensional system in the following section).

\subsection{Three-dimensional system}

\begin{figure}
\includegraphics[bb=20bp 20bp 592bp 480bp,scale=0.4]{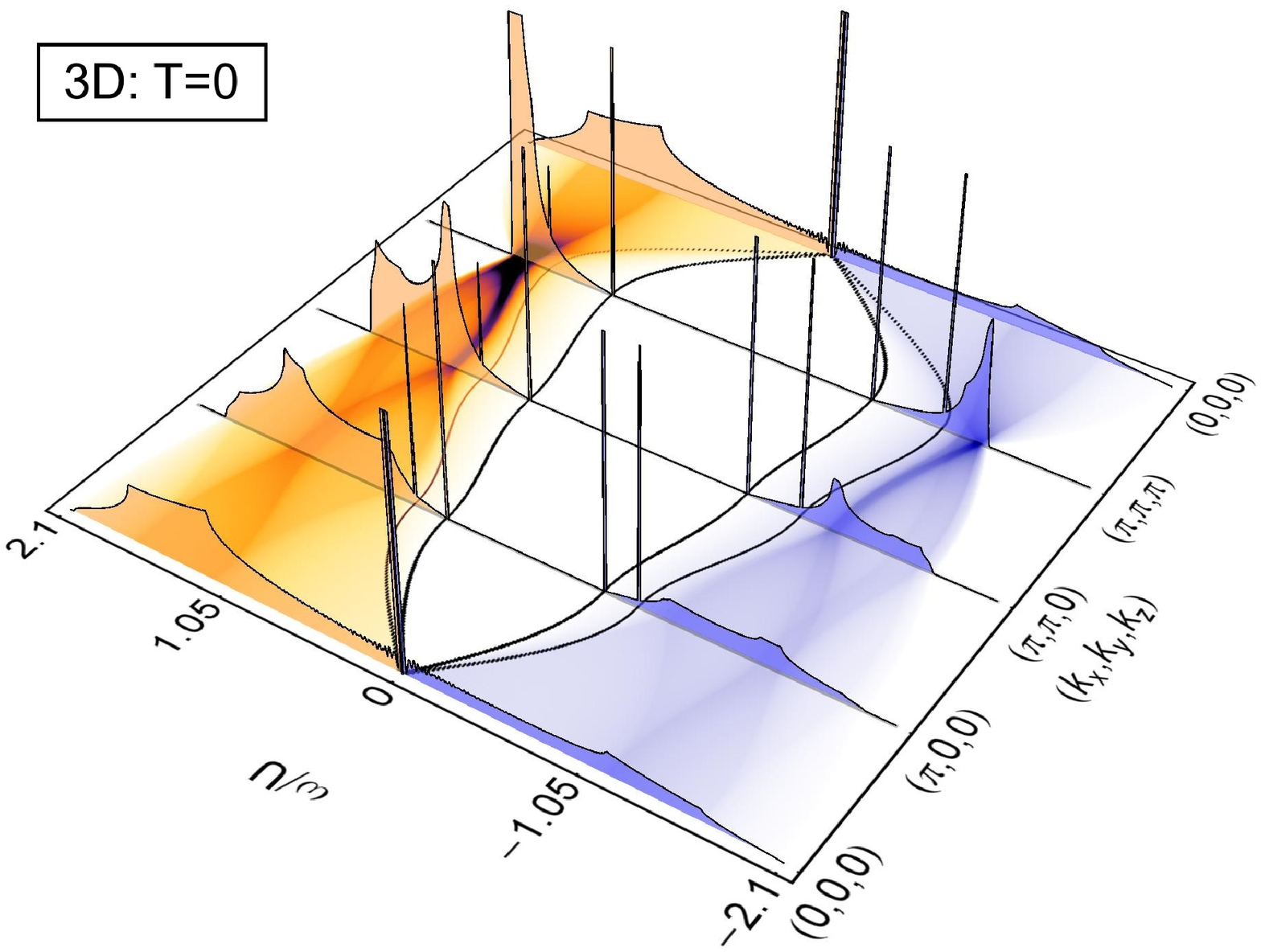}

\caption{(Color online) Density plot of the spectral function for three-dimensional
system at $k_{B}T/U=0$ and $t/U=0.045$, $\mu/U=0.5$ (darker areas
epics higher value) along with cuts of absolute value of $A\left(\mathbf{k}\omega\right)$
for selected values of momentum. For $\omega/U>0$, $A\left(\mathbf{k}\omega\right)>0$
(brown), while for $\omega/U<0$,\textbf{ $A\left(\mathbf{k}\omega\right)<0$}
(blue).\label{fig:3dT0}}
\end{figure}

\begin{figure}
\includegraphics[bb=20bp 20bp 592bp 480bp,scale=0.4]{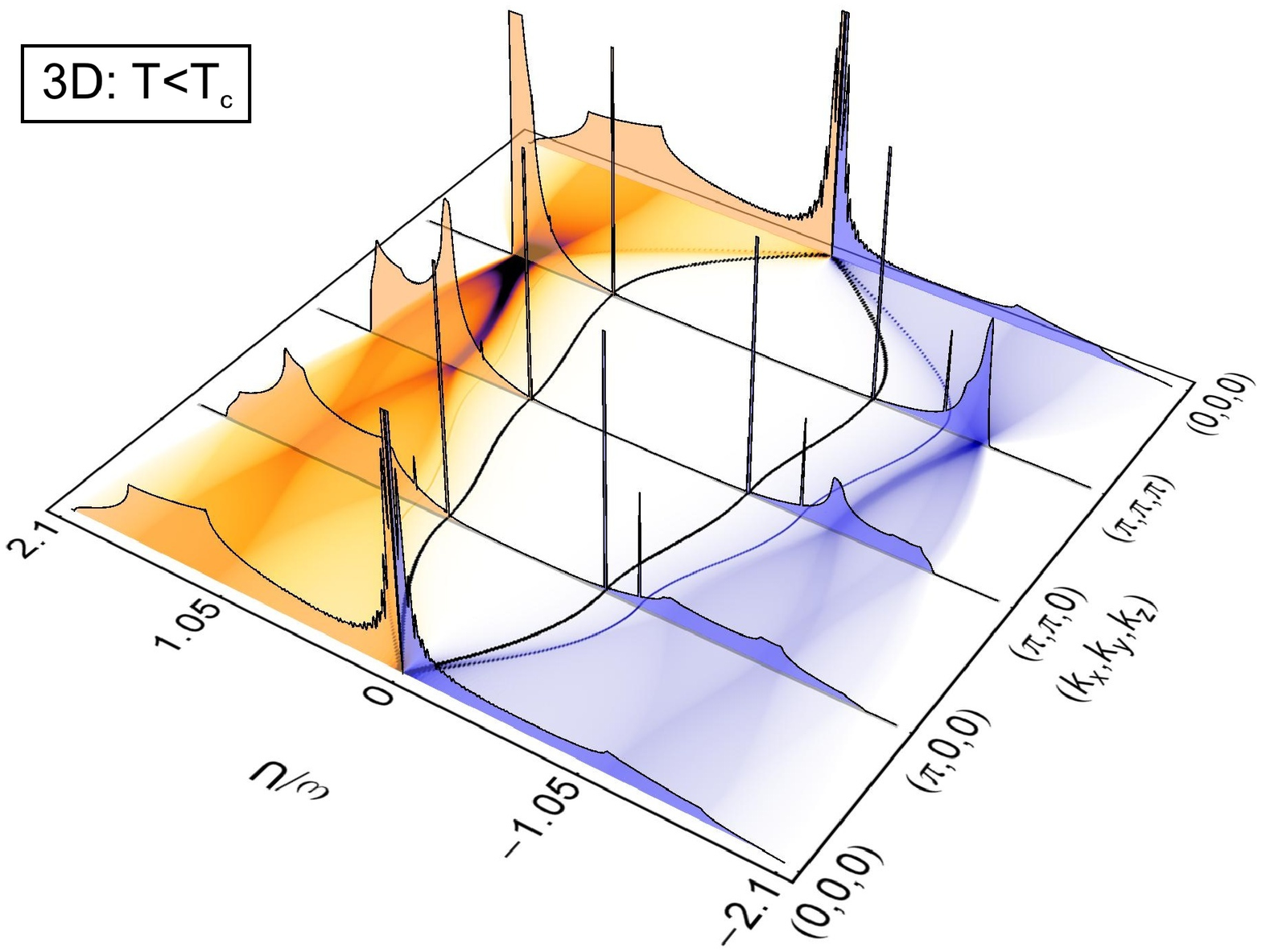}

\caption{(Color online) Density plot of the spectral function for three-dimensional
system at $k_{B}T/U=0.077$ (below critical temperature) and $t/U=0.045$,
$\mu/U=0.5$ (darker areas epics higher value) along with cuts of
absolute value of $A\left(\mathbf{k}\omega\right)$ for selected values
of momentum. For $\omega/U>0$, $A\left(\mathbf{k}\omega\right)>0$
(brown), while for $\omega/U<0$,\textbf{ $A\left(\mathbf{k}\omega\right)<0$}
(blue).\label{fig:3dT0_c}}
\end{figure}

\begin{figure}
\includegraphics[bb=20bp 20bp 592bp 480bp,scale=0.4]{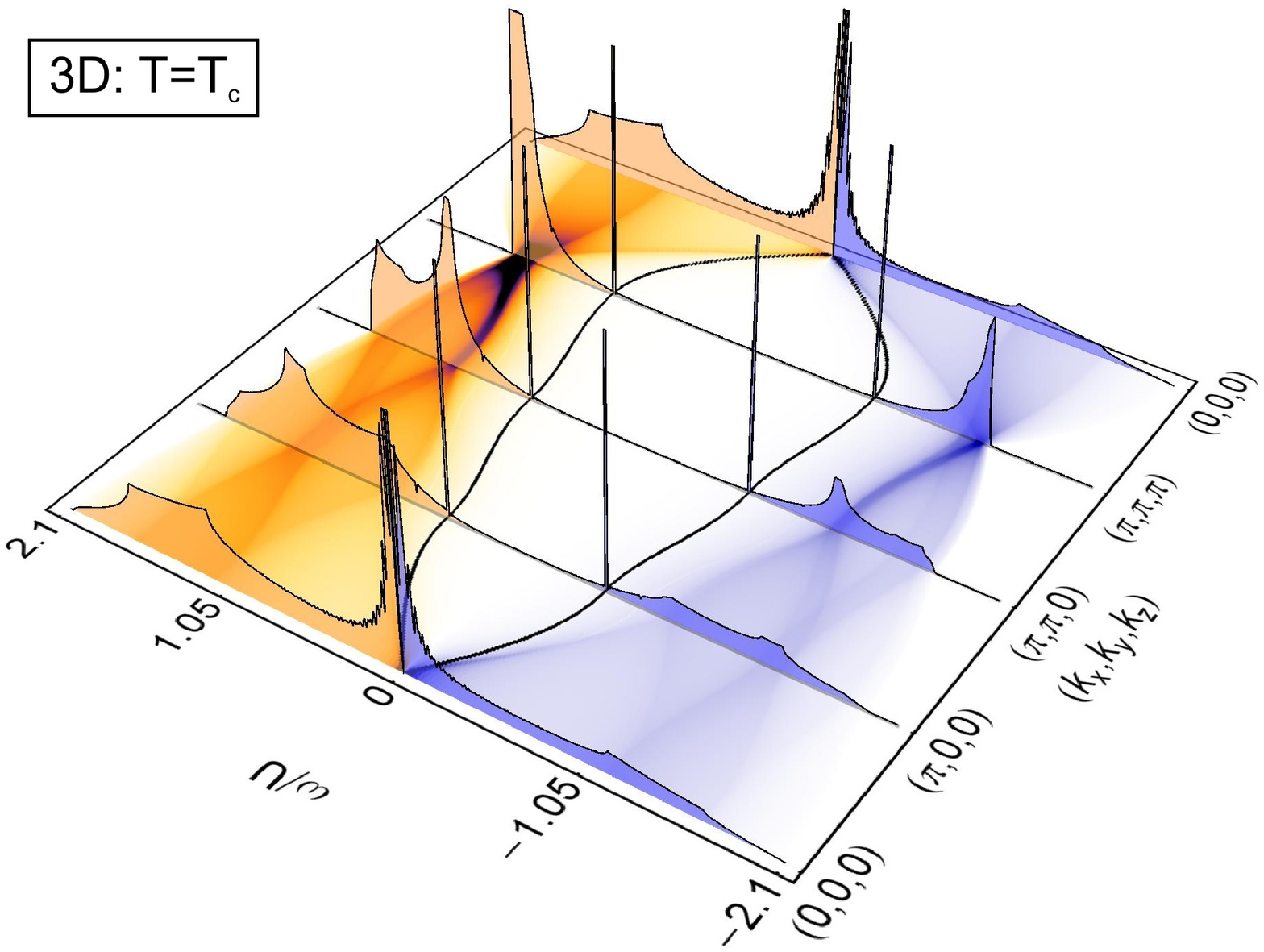}

\caption{(Color online) Density plot of the spectral function for three-dimensional
system at $k_{B}T/U=0.085$ (at the critical temperature) and $t/U=0.045$,
$\mu/U=0.5$ (darker areas epics higher value) along with cuts of
absolute value of $A\left(\mathbf{k}\omega\right)$ for selected values
of momentum. For $\omega/U>0$, $A\left(\mathbf{k}\omega\right)>0$
(brown), while for $\omega/U<0$,\textbf{ $A\left(\mathbf{k}\omega\right)<0$}
(blue).\label{fig:3dTc}}
\end{figure}

\begin{figure}
\includegraphics[bb=20bp 20bp 592bp 480bp,scale=0.4]{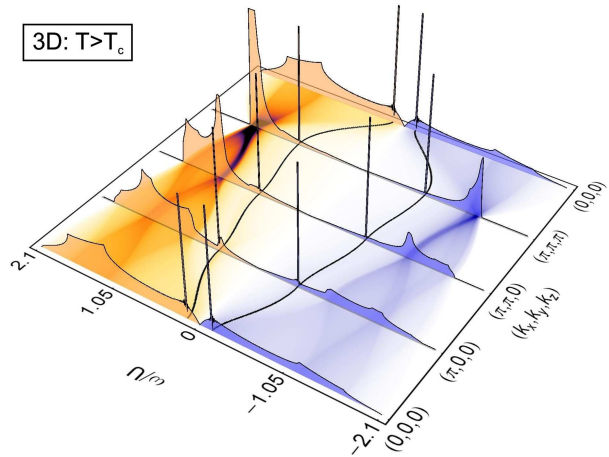}

\caption{(Color online) Density plot of the spectral function for three-dimensional
system at $k_{B}T/U=0.167$ (over the critical temperature) and $t/U=0.045$,
$\mu/U=0.5$ (darker areas epics higher value) along with cuts of
absolute value of $A\left(\mathbf{k}\omega\right)$ for selected values
of momentum. For $\omega/U>0$, $A\left(\mathbf{k}\omega\right)>0$
(brown), while for $\omega/U<0$,\textbf{ $A\left(\mathbf{k}\omega\right)<0$}
(blue).\label{fig:3dTh}}
\end{figure}

The evolution of the excitation spectra with increasing temperature
in the three-dimensional system is presented in Figs. \ref{fig:3dT0}-\ref{fig:3dTh}.
At $T=0$, for chosen values of $t/U=0.045$ and $\mu/U=0.5$, the
system is in the superfluid state, which is signaled by presence of
both coherence peaks: resulting from non-zero Bogoliubov amplitude
$b_{0}$ and long-range phase order ($\psi_{B}\ne0$). With the temperature
being raised, the weight of the phase order peak decreases (see, Fig.
\ref{fig:3dT0_c}), and at criticality it disappears completely (see,
Fig. \ref{fig:3dTc}). Above the critical temperature, a gap opens
at $\mathbf{k}=0$. However, it may become hidden, since thermal fluctuations
also smear the inner edges of the bands, thus filling the gap with
long tails of the decaying bands (see, Fig. \ref{fig:3dTh}).

\section{Conclusions}

The present paper extends our previous work (see, Ref. \cite{zaleski_spectral}),
which examined the excitation spectra at $T=0$ to finite temperatures.
For the bosonic system in optical lattice, as temperature increases,
thermal fluctuations melt away both the superfluid and Mott-insulating
phases, introducing the normal phase. We have determined the combined
effects of two and three dimensional lattice potential trapping and
temperature for a system of strongly interacting bosons on several
lattice structures. In order
to understand the properties of cold atoms in optical lattices, we
examined the dependence of momentum and energy-resolved excitation
spectra on the temperature. We showed that the general effect of thermal
fluctuations was twofold: first, the weight of the coherence peaks
resulting from the phase ordering being decreased with raising temperature.
Second, there was smearing of the inner edges of one-particle bands
tending to fill the excitation gap. However, in the two dimensional
system, in which the long-range ordered state exists only at zero
temperature, in the finite temperatures the sharp peak was recreated
(though this time it did not signal the superfluid phase). As a result,
the existence of the sharp peaks in the excitation spectra may not
necessarily be a reliable diagnostic tool of long-range order in these
systems. This is similar to results of Kato, et al. who showed that
sharp peaks in the time-of-flight experiments are not an unequivocal
proof of superfluidity \cite{triveditof}. On the other hand, in the
present paper, the three-dimensional system did not exhibit any remnants
of the phase-order coherence peak above the critical temperature.
This fact that may have important consequences for long range properties
in lower dimensions. The quantitative values for the excitation spectra
presented here provide benchmarks for continuing efforts to emulate
the Bose-Hubbard model on optical lattices, and demonstrate consistency
between strongly interacting atoms and phenomena observed in condensed-matter
systems.
\begin{acknowledgments}
We would like to acknowledge support from the Polish National Science
Centre (Grant No. 2011/03/B/ST3/00481).\end{acknowledgments}

\end{document}